\documentclass{article}

\makeatother
\usepackage{multicol,placeins}
\usepackage[preprint]{spconf}
\usepackage{graphicx}
\usepackage{tikz}
\usepackage{color}
\usepackage{amsmath, amsfonts, amssymb, mathrsfs}
\usepackage[amsmath,thmmarks]{ntheorem}
\usepackage{subfigure}
\usepackage{epstopdf}
\usepackage{pgfplots}
\usepackage{arydshln}

\usepackage[boxed]{algorithm2e}
\SetKwInput{KwParameter}{Parameters}

\makeatletter
\renewcommand*\env@matrix[1][*\c@MaxMatrixCols c]{%
  \hskip -\arraycolsep
  \let\@ifnextchar\new@ifnextchar
  \array{#1}} 

\theoremstyle{plain}
\theoremheaderfont{\itshape\bfseries}
\theorembodyfont{\normalfont\itshape}
\theoremseparator{:}
\theoremsymbol{}
\newtheorem{theorem}{Theorem}

\theoremstyle{nonumberplain}

\theoremstyle{nonumberplain}
\theoremheaderfont{\itshape}
\theorembodyfont{\normalfont}
\theoremseparator{:}
\theoremsymbol{}
\newtheorem{remark}{Remark}

\theoremstyle{plain}
\theoremheaderfont{\itshape\bfseries}
\theorembodyfont{\normalfont}
\theoremseparator{:}

\theoremstyle{nonumberplain}
\theoremsymbol{\rule{1ex}{1ex}}

\newlength\fheight
\newlength\fwidth
\def\d{ \mathrm{d} }						
\def\I{ \mathrm{i} }						
\def\E{ \mathrm{e} }						
\def\T{ \mathrm{T} }						
\def\NN{ \mathbb{N} }						
\def\ZN{ \mathbb{Z} }						
\def\RN{ \mathbb{R} }						
\def\CN{ \mathbb{C} }						
\def\B{ \mathcal{B} }						
\def\PW{ \mathcal{PW} }         
\def\ba{ \mathbf{a} }						
\def\be{ \mathbf{e} }						
\def\bx{ \mathbf{x} }
\def\by{ \mathbf{y} }
\def\bz{ \mathbf{z} }
\def\balpha{ \boldsymbol{\alpha} }
\def\bpsi{ \boldsymbol{\psi} }

\def\bA{ \mathbf{A} }
\def\bB{ \mathbf{B} }
\def\bX{ \mathbf{X} }
\def\bZ{ \mathbf{Z} }
\def\bPsi{ \mathbf{\Psi} }

\DeclareMathOperator*{\kr}{kr}			     
\DeclareMathOperator*{\argmin}{arg\,min}

\title{Compressive phase retrieval of sparse bandlimited signals}
\name{
  \c{C}a\u{g}kan~Yapar,
  Volker~Pohl,
	Holger~Boche\thanks{This work was partly supported by the German Research Foundation (DFG) under Grant BO~1734/20-1 and PO~1347/2-1.}
}
\address{
  Lehrstuhl f{\"u}r Theoretische Informationstechnik\\
  Technische Universit{\"a}t M{\"u}nchen, 80333 M{\"u}nchen, Germany\\
  \{cagkan.yapar, volker.pohl, boche\}@tum.de
}

\begin{document}
\ninept
\maketitle
%
\begin{abstract}
This contribution proposes a two stage strategy to allow for phase retrieval in state of the art sub-Nyquist sampling schemes for sparse multiband signals.
The proposed strategy is based on data acquisition via modulated wideband converters known from sub-Nyquist sampling.
This paper describes how the modulators have to be modified such that signal recovery from sub-Nyquist amplitude samples becomes possible and a corresponding recovery algorithm is given which is computational efficient.
In addition, the proposed strategy is fairly general, allowing for several constructions and recovery algorithms.
\end{abstract}

\section{Introduction}
\label{sec:intro}

In many applications it is not possible to measure the phase information of electromagnetic or acoustic signals. 
Then the phase information may be retrieved exploiting known signal characteristics.
For example, if the signal is known to be causal, the phase can be determined from the amplitude via Hilbert transform techniques \cite{Burge_PhaseProblem_1976}. Similarly, if the poles and zeros of the signal satisfies certain conditions then the phase can be determined from amplitude measurements \cite{Oppenheim_Phase_80}.
Alternatively, one can retrieve the phase from suitable chosen intensity measurements \cite{Fienup1982_PhaseRetrieval}.
In recent years, the phase retrieval problem for finite dimensional signal spaces attracted some interest \cite{Balan_RecWithoutPhase_06,Balan_Painless_09,CandesEldar_PhaseRetrieval,Bandeira_PR14,PYB_STIP14,bodmann2013stable}.
Later, these results were extended to sparse signals \cite{LiVoro_2012,Ohlsson_CPRL,Schnitter_prGAMP,Pedarsani_PhaseCode}.
In particular a two-stage technique for compressive phase retrieval was proposed in \cite{Pohl_ICASSP15,Iwen_2015} and applied, for example, in \cite{Bodmann_SAMPTA15}.
It basically allows the design of appropriated measurement procedures for the sparse phase retrieval problem by combining known algorithms and methods from (non-sparse) phase retrieval and compressive sensing.
Only little results exist for phase retrieval in infinite-dimensional signal spaces.
Results in \cite{Boche_Pohl_IEEE_IT08} indicate that Hilbert transform techniques for phase retrieval of causal signals can not be applied or only for sufficiently smooth signals \cite{PB_Book_AdvTopics}.
Signal recovery of real valued bandlimited signals from amplitude measurements taken at twice the Nyquist rate were considered in \cite{Thakur2011}.
Complex valued bandlimited signals were treated in \cite{Yang_SampTA13,PYB_JFAA14}, using a sampling rate of four times the Nyquist rate.

In this paper, we address the problem of reconstructing bandlimited signals with a sparsity prior from the amplitudes of sub-Nyquist samples.
Specifically, we assume a multiband structure where only the number of occupied bands is known but not the actual band locations.
It is shown how the two-stage technique from \cite{Pohl_ICASSP15,Iwen_2015} can be extended to sparse infinite dimensional signal spaces. 
In particular, we propose a two stage algorithm which combines the phase retrieval methodology proposed in \cite{Yang_SampTA13,PYB_JFAA14} with the sub-Nyquist sampling techniques from \cite{ME_BlindMultipand_IEEESP09,Tropp_SubNyquist10,ME_TheoryToPraxis10}.

\section{Signal Model and Notations}
\label{sec:signal}

\paragraph*{General notations}

As usual, $L^{p}(\RN)$, with $1\leq p\leq \infty$ stands for the spaces of Lebesgue integrable functions on the real axis $\RN$.
For any $x \in L^{1}(\RN)$ its \emph{Fourier transform} is defined by
\begin{eqnarray*}
	& \widehat{x}(f)
	= \big( \mathcal{F} x\big)(f)
	= \int_{\RN} x(t)\, \E^{-\I 2\pi f t}\, \d t\;,
	\quad f\in\RN\;.
\end{eqnarray*}
By Plancherel's theorem, $\mathcal{F}$ can be extended to a mapping \linebreak $\mathcal{F} : L^{2}(\RN) \to L^{2}(\RN)$ with $\|x\|_{2} = \|\widehat{x}\|_{2}$.

Vectors in the $N$-dimensional Euclidean vector space $\CN^{N}$ will be denoted by boldface lowercase letters, like $\bx = (x_{1},\dots,x_{N})^{\T}$, and matrices by boldface upper case letters, like $\bA$.
The \emph{Kruskal rank} $\kr(\bA)$ of a matrix  $\bA$ is the maximal $k\in\NN$ such that any $k$ columns of $\bA$ are linearly independent.
The set $\{\be_{n}\}^{N}_{n=1}$ will represent the canonical basis of the Cartesian coordinate system, i.e.
$\be_{1} = (1,0,0,\dots)^{\T}$, $\be_{2} = (0,1,0,\dots)^{\T}$, $\dots$.

\paragraph*{Signal model}
We consider signals with a multiband structure within the band-limits $[-f_{\mathrm{N}},f_{\mathrm{N}}]$ where $f_{\mathrm{N}} > 0$ is the \emph{Nyquist frequency} of the signals.
The frequency interval $[-f_{\mathrm{N}},f_{\mathrm{N}}]$ is divided into $L = 2L_{0}+1$ disjoint subbands $\mathcal{B}_{n}$ of length $\Omega := 2 f_{\mathrm{N}}/L$ of the form
\begin{equation*}
	\B_{n} = \big[ n \Omega - \tfrac{\Omega}{2} , n \Omega + \tfrac{\Omega}{2} \big]\;,
	\quad n=-L_{0},\dots,0,\dots,L_{0}
\end{equation*}
and where $L_{0}\in\NN$.
To simplify the presentation, but without loss of generality, it is assumed that $L$ is an odd number.
It is assumed that the signals are supported on only $N \ll L$ frequency bands.
The indices of the occupied bands are collected in the set $\mathcal{I} \subset \{-L_{0},\dots,L_{0}\}$ with cardinality $|\mathcal{I}| = N$.
Then $\B = \cup_{n\in\mathcal{I}} \B_{n}$ is the frequency support of our signals,
and $L^{2}(\B)$ is the set of all square integrable functions on $\B$.
We write 
\begin{eqnarray*}
	& \PW_{\B} = \Big\{ x(t) = \int_{\B} \widehat{x}(f)\, \E^{\I 2\pi f t}\, \d f\ :\ \widehat{x} \in L^{2}(\B) \Big\}
\end{eqnarray*}
for the set of all square integrable functions which can be represented as the inverse Fourier transform of a function in $L^{2}(\B)$ equipped with the usual $L^{2}(\RN)$ norm. It is assumed that we know the number of occupied bands $N$ but not the band locations $\mathcal{N}$.

So every $x \in \PW_{\B}$ can be written as the sum $L$ mutually orthogonal signals $\widetilde{x}_{l}$, each of which is supported on one of the frequency intervals $\B_{l}$ of length $\Omega$.
Thus
$x = \sum_{l\in\mathcal{N}} \widetilde{x}_{l}$ with
$\widetilde{x}_{l} \in \PW_{\B_{l}}$.
Later, we denote by $x_{l}(t) = \widetilde{x}(t)\, \E^{\I 2\pi l \Omega t}$ the signal component of $l$-th frequency band, shifted to the frequency interval $[-\Omega/2,\Omega/2]$.

\section{Measurement Setup}
\label{sec:demod}

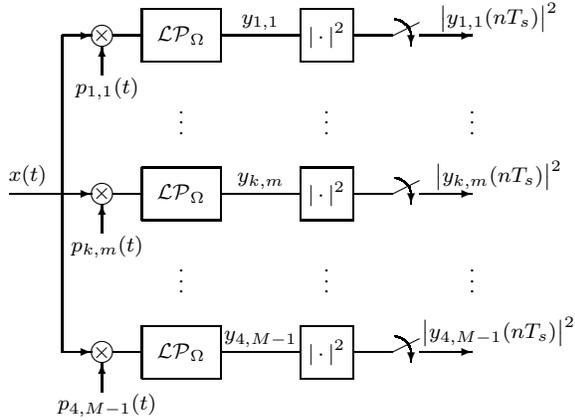
\begin{figure}[t]
\begin{center}
\begin{picture}(200,145)(20,25)
	\put(27,107){\makebox(0,0){\footnotesize $x(t)$}}
	\put(20,100){\line(1,0){20}}
	\put(40,40){\line(0,1){120}}
	\put(40,160){\vector(1,0){11}}
	\put(55,160){\circle{8}}		\put(55,160){\makebox(0,0){\footnotesize$\times$}}
	\put(55,145){\vector(0,1){11}}	\put(57,140){\makebox(0,0){\footnotesize $p_{1,1}(t)$}}
	\put(59,160){\line(1,0){11}}
	\put(40,100){\vector(1,0){11}}
	\put(55,100){\circle{8}}		\put(55,100){\makebox(0,0){\footnotesize$\times$}}
	\put(55,85){\vector(0,1){11}}	\put(57,80){\makebox(0,0){\footnotesize $p_{k,m}(t)$}}
	\put(59,100){\line(1,0){11}}
	\put(40,40){\vector(1,0){11}}
	\put(55,40){\circle{8}}   	\put(55,40){\makebox(0,0){\footnotesize$\times$}}
	\put(55,25){\vector(0,1){11}}	\put(57,20){\makebox(0,0){\footnotesize $p_{4,M-1}(t)$}}
	\put(59,40){\line(1,0){11}}
	\put(70,170){\line(1,0){30}}
	\put(70,150){\line(1,0){30}}
	\put(70,150){\line(0,1){20}}
	\put(100,150){\line(0,1){20}}
	\put(85,160){\makebox(0,0){\footnotesize $\mathcal{LP}_{\Omega}$}}
	\put(100,160){\line(1,0){30}}
	\put(115,165){\makebox(0,0){\footnotesize $y_{1,1}$}}
	\put(130,170){\line(1,0){20}}
	\put(130,150){\line(1,0){20}}
	\put(130,150){\line(0,1){20}}
	\put(150,150){\line(0,1){20}}
	\put(140,160){\makebox(0,0){\footnotesize $\left|\, \cdot\, \right|^{2}$}}
	\put(150,160){\line(1,0){15}}	
	\put(85,130){\makebox(0,0){\footnotesize $\vdots$}}
	\put(140,130){\makebox(0,0){\footnotesize $\vdots$}}
	\put(195,130){\makebox(0,0){\footnotesize $\vdots$}}			
	\put(70,110){\line(1,0){30}}
	\put(70,90){\line(1,0){30}}
	\put(70,90){\line(0,1){20}}
	\put(100,90){\line(0,1){20}}
	\put(85,100){\makebox(0,0){\footnotesize $\mathcal{LP}_{\Omega}$}}
	\put(100,100){\line(1,0){30}}
	\put(115,105){\makebox(0,0){\footnotesize $y_{k,m}$}}
	\put(130,110){\line(1,0){20}}
	\put(130,90){\line(1,0){20}}
	\put(130,90){\line(0,1){20}}
	\put(150,90){\line(0,1){20}}
	\put(140,100){\makebox(0,0){\footnotesize $\left|\, \cdot\, \right|^{2}$}}
	\put(150,100){\line(1,0){15}}	
	\put(85,70){\makebox(0,0){\footnotesize $\vdots$}}
	\put(140,70){\makebox(0,0){\footnotesize $\vdots$}}	
	\put(195,70){\makebox(0,0){\footnotesize $\vdots$}}		
	\put(70,50){\line(1,0){30}}
	\put(70,30){\line(1,0){30}}
	\put(70,30){\line(0,1){20}}
	\put(100,30){\line(0,1){20}}
	\put(85,40){\makebox(0,0){\footnotesize $\mathcal{LP}_{\Omega}$}}
	\put(100,40){\line(1,0){30}}
	\put(115,45){\makebox(0,0){\footnotesize $y_{4,M-1}$}}
  \put(130,50){\line(1,0){20}}
	\put(130,30){\line(1,0){20}}
	\put(130,30){\line(0,1){20}}
	\put(150,30){\line(0,1){20}}
	\put(140,40){\makebox(0,0){\footnotesize $\left|\, \cdot\, \right|^{2}$}}
	\put(150,40){\line(1,0){15}}
	\put(165,160){\line(2,1){10}}
	\qbezier(166,166)(171,165)(172,160)	\put(172,158){\vector(0,-1){2}}
	\put(175,160){\vector(1,0){20}}
	\put(205,167){\makebox(0,0){\footnotesize $\big| y_{1,1}(n T_{s}) \big|^{2}$}}
	\put(165,100){\line(2,1){10}}
	\qbezier(166,106)(171,105)(172,100)	\put(172,98){\vector(0,-1){2}}
	\put(175,100){\vector(1,0){20}}
	\put(205,107){\makebox(0,0){\footnotesize $\big| y_{k,m}(n T_{s}) \big|^{2}$}}	
	\put(165,40){\line(2,1){10}}
	\qbezier(166,46)(171,45)(172,40)		\put(172,38){\vector(0,-1){2}}
	\put(175,40){\vector(1,0){20}}
	\put(205,48){\makebox(0,0){\footnotesize $\big| y_{4,M-1}(n T_{s}) \big|^{2}$}}
\end{picture}
\end{center}
\caption{Modulated wideband converter with $4(M-1)$ mixing branches.
Therein $\mathcal{LP}_{\Omega}$ symbolizes an ideal low-pass filter with cutoff $\Omega/2$, $|\cdot|^{2}$ stands for squared amplitude measurements. The mixing sequences $p_{k,m}$ are defined in \eqref{equ:MixSeq}, and the sampling interval is $T_{s} = 1/\Omega$. }
\label{fig:Modul1}
\end{figure}

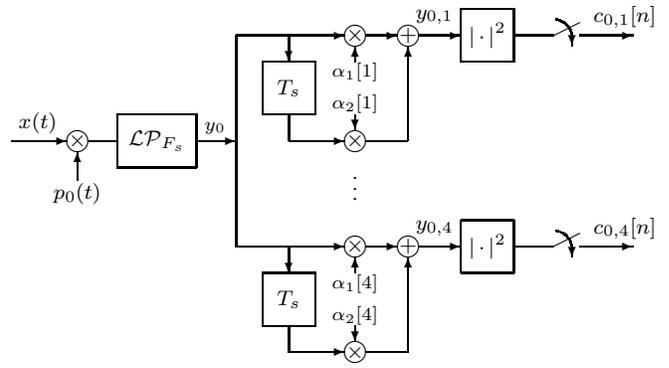
\begin{figure}[t]
\begin{center}
\begin{picture}(230,130)(0,40)
	\put(0,127){\makebox(0,0){\footnotesize $x(t)$}}
	\put(-10,120){\vector(1,0){21}}
	\put(15,120){\circle{8}}		\put(15,120){\makebox(0,0){\footnotesize$\times$}}
	\put(15,105){\vector(0,1){11}}	\put(15,100){\makebox(0,0){\footnotesize $p_{0}(t)$}}
	\put(19,120){\line(1,0){11}}
	\put(30,130){\line(1,0){30}}
	\put(30,110){\line(1,0){30}}
	\put(30,110){\line(0,1){20}}
	\put(60,110){\line(0,1){20}}
	\put(45,120){\makebox(0,0){\footnotesize $\mathcal{LP}_{F_{s}}$}}
	\put(67,125){\makebox(0,0){\scriptsize $y_{0}$}}
	\put(60,120){\vector(1,0){15}}
	\put(75,80){\line(0,1){80}}
	\put(75,160){\vector(1,0){41}}
	\put(120,160){\circle{8}}		\put(120,160){\makebox(0,0){\footnotesize$\times$}}
	\put(120,150){\vector(0,1){6}}
	\put(120,146){\makebox(0,0){\scriptsize $\alpha_{1}[1]$}}
	\put(120,120){\circle{8}}		\put(120,120){\makebox(0,0){\footnotesize$\times$}}
	\put(120,130){\vector(0,-1){6}}
	\put(120,134){\makebox(0,0){\scriptsize $\alpha_{2}[1]$}}

	\put(95,160){\vector(0,-1){10}}
	\put(85,150){\line(1,0){20}}
	\put(85,130){\line(1,0){20}}
	\put(85,130){\line(0,1){20}}
	\put(105,130){\line(0,1){20}}
	\put(95,140){\makebox(0,0){\footnotesize $T_{s}$}}
	\put(95,130){\line(0,-1){10}}
	\put(95,120){\vector(1,0){21}}
	\put(124,160){\vector(1,0){12}}
	\put(124,120){\line(1,0){16}}
	\put(140,120){\vector(0,1){36}}
	\put(140,160){\circle{8}}		\put(140,160){\makebox(0,0){\footnotesize$+$}}
	\put(150,167){\makebox(0,0){\footnotesize $y_{0,1}$}}
	\put(144,160){\vector(1,0){16}}
	\put(160,150){\line(1,0){20}}
	\put(160,170){\line(1,0){20}}
	\put(160,150){\line(0,1){20}}
	\put(180,150){\line(0,1){20}}
	\put(170,160){\makebox(0,0){\footnotesize $\left|\, \cdot\, \right|^{2}$}}
	\put(180,160){\line(1,0){15}}
	\put(195,160){\line(2,1){10}}
	\qbezier(196,166)(201,165)(202,160)	\put(202,158){\vector(0,-1){2}}
	\put(205,160){\vector(1,0){20}}
	\put(222,167){\makebox(0,0){\footnotesize $c_{0,1}[n]$}}
	\put(120,105){\makebox(0,0){\footnotesize $\vdots$}}
	\put(75,80){\vector(1,0){41}}
	\put(120,80){\circle{8}}		\put(120,80){\makebox(0,0){\footnotesize$\times$}}
	\put(120,70){\vector(0,1){6}}
	\put(120,66){\makebox(0,0){\scriptsize $\alpha_{1}[4]$}}
	\put(120,40){\circle{8}}		\put(120,40){\makebox(0,0){\footnotesize$\times$}}
	\put(120,50){\vector(0,-1){6}}
	\put(120,54){\makebox(0,0){\scriptsize $\alpha_{2}[4]$}}
	\put(95,80){\vector(0,-1){10}}
	\put(85,70){\line(1,0){20}}
	\put(85,50){\line(1,0){20}}
	\put(85,50){\line(0,1){20}}
	\put(105,50){\line(0,1){20}}
	\put(95,60){\makebox(0,0){\footnotesize $T_{s}$}}
	\put(95,50){\line(0,-1){10}}
	\put(95,40){\vector(1,0){21}}
	\put(124,80){\vector(1,0){12}}
	\put(124,40){\line(1,0){16}}
	\put(140,40){\vector(0,1){36}}
	\put(140,80){\circle{8}}		\put(140,80){\makebox(0,0){\footnotesize$+$}}
	\put(150,87){\makebox(0,0){\footnotesize $y_{0,4}$}}
	\put(144,80){\vector(1,0){16}}
	\put(160,70){\line(1,0){20}}
	\put(160,90){\line(1,0){20}}
	\put(160,70){\line(0,1){20}}
	\put(180,70){\line(0,1){20}}
	\put(170,80){\makebox(0,0){\footnotesize $\left|\, \cdot\, \right|^{2}$}}
	\put(180,80){\line(1,0){15}}
	\put(195,80){\line(2,1){10}}
	\qbezier(196,86)(201,85)(202,80)	\put(202,78){\vector(0,-1){2}}
	\put(205,80){\vector(1,0){20}}
	\put(222,87){\makebox(0,0){\footnotesize $c_{0,4}[n]$}}
\end{picture}
\end{center}
\caption{Preprocessing for phase propagation.}
\label{fig:Modul2}
\end{figure}

For blind signal recovery of multiband signals, the authors in \cite{Tropp_SubNyquist10,ME_TheoryToPraxis10} proposed the use of modulators (cf. Fig.~\ref{fig:Modul1}) to facilitate signal recovery from sub-Nyquist samples.
We adopt a similar approach but we modify the modulating functions $p_{k,m}$ to allow for signal recovery from amplitude samples only.
However, since no phase information is available, we will need an additional pre-processing step, as sketched in Fig.~\ref{fig:Modul2}.
Both setups are explained next.

\paragraph*{Modulated wideband converter}
The main measurement setup is shown in Fig.~\ref{fig:Modul1}. It has a similar structure as the modulated wideband converter in \cite{ME_TheoryToPraxis10}:
The signal $x \in \PW_{\B}$ enters $4 (M-1)$ channels. In each channel, the signal is multiplied by a modulating function $p_{k,m}$.
These functions are $T_{p} = 1/\Omega$ periodic and have the following form 
\begin{eqnarray}
\label{equ:MixSeq}
	& p_{k,m}(t) = \sum^{L_0}_{l=-L_0} a_{k,m}[l]\, \E^{\I \frac{2\pi}{T_{p}} l t}\,,
	\quad
	\begin{array}{l}
	k = 1,\dots,4\\
	m = 1,\dots,M-1
	\end{array}\;.
\end{eqnarray}
The coefficients $a_{k,m}[l] \in \CN$ will be determined subsequently.
After the modulators, the signals $x_{k,m}(t) = x(t)\, p_{k,m}(t)$ are filtered by an ideal low-pass filter with cutoff frequency $\Omega/2$.
This yields signals $y_{k,m}$ which are given in the frequency domain by
\begin{equation}
\label{equ:MixedSig}
	\widehat{y}_{k,m}(f) = \sum^{L_{0}}_{l=-L_{0}} a_{k,m}[l]\, \widehat{x}\big(f + l\Omega \big)\;,
	\quad f\in [-\Omega/2,\Omega/2]\;,
\end{equation}
using that $\Omega = 1/T_{p}$.
So each $y_{k,m}$ is bandlimited with bandwith $\Omega/2$.
Finally, we sample the squared amplitude of the signals $y_{k,m}$ at Nyquist rate $\Omega = 1/T_{s}$.
This yields the samples
\begin{equation}
\label{equ:MainSamples}
	|y_{k,m}[n]|^{2} = |y_{k,m}(n\, T_{s})|^{2}\;,
	\qquad n\in\ZN
\end{equation}
in each branch $k=1,\dots,4$ and $m=1,\dots,M-1$.

To simplify notations and for the subsequent derivations, we write all equations \eqref{equ:MixedSig} in matrix form as
\begin{equation}
\label{equ:SamplingFDomain}
	\widehat{\by}(f) = \bA\, \widehat{\bx}(f)\;,
	\quad f\in [-\Omega/2,\Omega/2]
\end{equation}
where $\widehat{\by}(f)$ is a vector of length $4 M - 4$ composed of the functions $\widehat{y}_{k,m}(f)$.
The vector $\widehat{\bx}(f)$ has length $L = 2 L_{0} + 1$ and contains the functions $\widehat{x}_{l}(f) := \widehat{x}(f + l\Omega)$.
The rows of the $(4 M - 4)\times L$ matrix $\bA$ contain the coefficients $a_{k,m}$ of all modulating functions.
Since $y_{k,m}$ is bandlimited, we can write $\widehat{y}_{k,m}$ as a Fourier series of the time domain samples
\begin{equation*}
	\widehat{y}_{k,m}(f)
	= \frac{1}{\Omega} \sum_{n\in\ZN} y_{k,m}[n]\, \E^{- \I \frac{2\pi}{\Omega}n f}\;,
	\quad f\in [-\Omega/2,\Omega/2]\;, 
\end{equation*}
and a completely similar relation holds for $\widehat{x}_{l}$.
Therefore \eqref{equ:SamplingFDomain} can be rewritten in the time domain as
\begin{equation}
\label{equ:MatrixSamples}
	\by[n] = \bA\, \bx[n]\;,
	\qquad n\in\ZN\;,
\end{equation}
where the vectors $\by[n]$ and $\bx[n]$ contain the samples $y_{k,m}[n] = y_{k,m}(n T_{s})$ and $x_{l}[n] = x_{l}(n T_{s})$, respectively.
So the $l$-th entry of the vector $\bx[n]$ is equivalent to the samples of the signal component which is supported on the $l$-th frequency band.

\paragraph*{Modulator for phase propagation}
Additionally to the modulated wideband converter described in the previous paragraph,
we apply another preprocessing step, sketched in Fig~\ref{fig:Modul2}.
Similar as in Fig.~\ref{fig:Modul1}, the signal $x\in\PW_{\B}$ is modulated by a $T_{p}$-periodic function
\begin{eqnarray*}
	& p_{0}(t) = \sum^{L_0}_{l=-L_0} a_{0}[l]\, \E^{\I \frac{2\pi}{T_{p}} l t}
\end{eqnarray*}
and filtered by an ideal low-pass with cutoff frequency $\Omega/2$.
This yields the signal $y_{0}(t) = x(t)\, p_{0}(t)$ with Fourier transform
\begin{eqnarray}
\label{equ:Mod2}
	& \widehat{y}_{0}(f) = \sum^{L_{0}}_{l=-L_{0}} a_{0}[l]\, \widehat{x}\big(f + l\Omega \big)\;,
	\quad f\in [-\Omega/2,\Omega/2]\;.
\end{eqnarray}
Again, we define $\widehat{x}_{l}(f) := \widehat{x}(f + l\Omega)$, and
since $y_{0}$ and $x_{l}$ are bandlimited to $[-\Omega/2,\Omega/2]$, we can rewrite \eqref{equ:Mod2} equivalently in terms of the samples of $y_{0}$ and $x_{l}$ in the time domain:
\begin{eqnarray*}
	& y_{0}[n]
	:= y_{0}(n\T_{s})
	= \sum^{L_{0}}_{l=-L_{0}} a_{0}[l]\, x_{l}[n]
	= \ba_{0}^{\T}\, \bx[n]\;,
	\quad n\in\ZN\;,
\end{eqnarray*}
where $\ba_{0}^{\T}$ is a row vector containing the coefficients $\{ a_{0}[l] \}_{l=-L_{0}}^{L_{0}}$.
Then the modulated signal $y_{0}$ is delayed by a sampling period $T_{s}$.
Afterwards, we have four branches ($k = 1,\dots,4$) in which the undelayed and the delayed signals are linearly combined with coefficients $\alpha_{1}[k]$ and $\alpha_{2}[k]$, respectively. This yields $y_{k,0}(t) = \alpha_{1}[k]\, y_{0}(t) + \alpha_{2}[k]\, y_{0}(t-T_{s})$.
Finally, the squared modulus of these signals are sampled at rate $R_{s} = 1/T_{s} = \Omega$.
This gives four  measurement sequences ($k=1,\dots,4$) which can be written as
\begin{align}
\label{equ:Samples2}
	c_{k}[n]
	&= \left| y_{k,0}(n  T_{s}) \right|^{2}
	= \left| \overline{\alpha_{1}[k]}\, y_{0}(n T_{s}) + \overline{\alpha_{2}[k]}\, y_{0}([n-1] T_{s}) \right|^{2}\nonumber\\
	&= \left|\left\langle \by_{0}[n] , \balpha[k] \right\rangle_{\CN^{2}}\right|^{2}\;,
	\qquad n\in\ZN
\end{align}
with the $\CN^{2}$-vectors
\begin{equation*}
	\balpha[k] = \binom{\alpha_{1}[k]}{\alpha_{2}[k]}
	\quad\text{and}\quad
	\by_{0}[n] = \binom{y_{0}(nT_{s})}{y_{0}([n-1] T_{s})}\;.
\end{equation*}

\paragraph*{Specification of parameters}

It remains to specify the coefficients of the modulating functions $p_{k,m}$ and $p_{0}$, and of the coefficient vectors $\balpha[k]$.
The coefficients of all functions $p_{k,m}$ are collected in the matrix $\bA$ in \eqref{equ:MatrixSamples}.
To specify this $(4M-4) \times L$ matrix, we follow the ideas in \cite{Pohl_ICASSP15,Iwen_2015} and write $\bA$ as a product $\bA = \bPsi\, \bB$.
Therein $\bPsi$ is a matrix of size $(4M-4) \times M$ which will be chosen such that phase retrieval in $\CN^{M}$ is possible.
The $M\times L$ matrix $\bB$ is chosen to be good compressive sampling measurement matrix.
With this factorization of $\bA$, we rewrite \eqref{equ:MatrixSamples} as
\begin{equation}
\label{equ:Samples1b}
	\by[n] = \bPsi\, \bz[n]
	\qquad\text{with}\qquad
	\bz[n] := \bB\, \bx[n]\;.
\end{equation}

We choose the phase retrieval matrix $\bPsi$ according to the phase retrieval approach in \cite{PYB_STIP14}.
To this end, we define the following elementary vectors in $\bpsi_{k,m} \in \CN^{M}$
\begin{equation}
\label{equ:MeasVect}
\begin{array}{rclcrcl}
	\bpsi_{1,m} & = & \phantom{-} a\, \be_{1} + b\, \be_{m+1}\\[0.5ex]
	\bpsi_{2,m} & = & \phantom{-} b\, \be_{1} + a\, \be_{m+1}\\[0.5ex]
	\bpsi_{3,m} & = & \phantom{-} a\, \be_{1} - b\, \be_{m+1}\\[0.5ex]
	\bpsi_{4,m} & = & -b\, \be_{1} + a\, \be_{m+1}
\end{array}\,,
\qquad m=1,2,\dots,M-1
\end{equation}
with the two constants 
\begin{equation}
\label{equ:constants}
	a = \sqrt{\tfrac{1}{2}\big(1-\tfrac{1}{\sqrt{3}}\big)}
	\quad\text{and}\quad
	b = \E^{\I 5\pi/4}\sqrt{\tfrac{1}{2}\big(1+\tfrac{1}{\sqrt{3}} \big)}\;.
\end{equation}
Then the $n$-th row of $\bPsi$ is equal to $\bpsi^{*}_{m,k}$ with $n=4(m-1)+k$.

To specify the compressive sampling matrix $\bB$, we notice that $\bx[n] \in \CN^{L}$ is an $N$-sparse vector for every $n\in\ZN$.
We will see later that we have to solve the right hand side equation of \eqref{equ:Samples1b} for signal recovery.
Consequently, we choose $\bB$ such that this equation has a unique $N$-sparse solution for every $\bz[n] \in \CN^{M}$.
A necessary and sufficient condition for that is $\kr(\bB) \geq 2 N$, which requires $M \geq 2N$.
Moreover, there exist several specific constructions (see, e.g., \cite[Chap.~2.2]{FoucartRauhut_CS}) of matrices $\bB \in \CN^{2N\times L}$ with $\kr(\bB) = 2N$.
So we assume in the following that $\bB$ is chosen such that $\kr(\bB) \geq 2 N$.

To fix the setup in Fig.~\ref{fig:Modul2}, we specify the row vector $\ba_{0}^{\T}$, which contains the coefficients of the modulation function $p_{0}$, to be equal to the first row of the compressive sampling matrix $\bB$.
Moreover, the constants $\balpha[k] = (\alpha_{1}[k] , \alpha_{2}[k] )^{\T} \subset \CN^{2}$ are set to
\begin{equation}
\label{equ:VecAlpha}
	\balpha[1] =  \binom{a}{b},\
	\balpha[2] =  \binom{b}{a},\
	\balpha[3] =  \binom{a}{-b},\
	\balpha[4] =  \binom{-b}{a}\;.
\end{equation}
with constants $a,b \in \CN$ given in \eqref{equ:constants}.
By this choice of parameters, the samples in Fig.~\ref{fig:Modul2} are given as in \eqref{equ:Samples2} but with
\begin{equation}
\label{equ:by0}
	\by_{0}[n] = \binom{z_{1}[n]}{z_{1}[n-1]}
\end{equation}
where $z_{1}[n]$ is the first entry of the vectors $\bz[n]$ defined in \eqref{equ:Samples1b}.

\section{Recovery Procedure}
\label{sec:Recov}

At the output of the sampling system in Fig.~\ref{fig:Modul1}, one obtains the sequence of squared amplitude samples $\{|\by[n]|^{2}\}_{n\in\ZN}$.
From these samples, we want to reconstruct the original signal $x \in \PW_{\B}$.
In view of \eqref{equ:Samples1b}, signal recovery follows several steps.
First for every $n\in\ZN$, we determine $\bz[n] \in \CN^{M}$ from the amplitude measurements $|\by[n]|^{2}$ using a phase retrieval algorithm.
Second, we reconstruct $\bx[n]\in\CN^{L}$ from $\bz[n]$ using ideas of compressive sampling. 
Finally, $x$ can be reconstructed from the sequence $\{ \bx[n] \}_{n\in\ZN}$ by applying Shannon's sampling series.

\paragraph*{I. Phase retrieval step} 
The left hand side equation of \eqref{equ:Samples1b} shows that for every fixed sampling instant $n\in\ZN$, the amplitude measurements \eqref{equ:MainSamples} can be written as
\begin{equation}
\label{equ:QuadrMeasur}
	c_{k,m} = |y_{k,m}[n]|^{2}
	= \Big| \left\langle \bz[n] , \bpsi_{k,m}\right\rangle_{\CN^{M}} \Big|^{2},
	\
	\begin{array}{l}
	k=1,\dots,4\\
	m=1,\dots,M-1\;.
	\end{array}
\end{equation}
This is a usual phase retrieval problem for the unknown vector $\bz[n] \in \CN^{M}$ and with $4M-4$ measurement vectors $\bpsi_{m,k}$.
It was shown in \cite{PYB_STIP14} that the system of measurement vectors \eqref{equ:MeasVect} allows to determine $\bz[n]$ from the quadratic measurements \eqref{equ:QuadrMeasur} as long as $z_{1}[n] \neq 0$. Then an efficient algebraic recovery algorithm is described in \cite{PYB_STIP14}.
However, the quadratic measurements \eqref{equ:QuadrMeasur} determine $\bz[n]$ uniquely only up to an unknown phase factor.
So for every $n\in\ZN$, we are able determine $\widetilde{\bz}[n] = \bz[n]\, \E^{\I\theta_{n}}$ with unknown phases $\{\theta_{n}\}_{n\in\ZN}$.

\paragraph*{II. Phase propagation}
This step determines the unknown phases $\{\theta_{n}\}_{n\in\ZN}$.
To this end, we use the samples acquired with the setup in Fig.~\ref{fig:Modul2}.
The unknown phases are determined successively from these measurements as follows:
We begin, by setting the first phase (say $\theta_{0})$ arbitrarily, e.g. $\theta_{0} = 0$.
Assume that we already determined $\theta_{n-1}$.
At sampling instant $n$, we obtain the four quadratic measurements $\{c_{k}[n]\}^{4}_{k=1}$ given in \eqref{equ:Samples2} for the unknown vector \eqref{equ:by0}.
Again, this is a usual phase retrieval problem in $\CN^{2}$ with measurement vectors $\{\balpha[k]\}^{4}_{k=1}$.
These four vectors are designed according to the methodology from \cite{Balan_Painless_09} which allows an efficient recovery of 
$\by_{0}[n] = ( z_{1}[n],z_{1}[n-1])^{\T}$ up to a constant phase factor.
From this result, we can determine the relative phase difference $\Delta\theta_{n}$ between $z_{1}[n]$ and its predecessor $z_{1}[n-1]$.
Since $z_{1}[n]$ is the first entry in the vector $\bz[n]$, we can use $\Delta\theta_{n}$ to determine the unknown phase $\theta_{n}$ from $\theta_{n-1}$.
In this way, we are able to determine successively all $\theta_{n}$, except $\theta_{0}$.
So at the end of this step, we have determined the sequence $\{\bz[n]\}_{n\in\ZN}$ up to a global unknown phase factor $\theta_{0}$.

\paragraph*{III. CS based signal recovery}

Next we determine $\bx[n]$ from the previously determined $\bz[n]$ by solving the linear system of equations on the right hand side of \eqref{equ:Samples1b}.
The vector $\bx[n] \in \CN^{L}$ contains samples of the signal in the $l$-th frequency band.
Since only $N\ll L$ signal bands are occupied, the vector $\bx[n]$ is $N$-sparse.
So we can solve \eqref{equ:Samples1b} formally for every $n\in\ZN$ by
\begin{equation*}
	\argmin_{\bx[n] \in\CN^{L}} \big\| \bx[n]\big\|_{0}
	\quad\text{s.t.}\quad
	\bz[n] = \bB\, \bx[n]\;.
\end{equation*}
This optimization problem has a unique $N$-sparse solution because $\bB$ was chosen such that $\kr(\bB) \geq 2N$.
However, it is also known that this minimization problem is NP-hard.
Nevertheless, other standard CS recovery methods (like basis pursuit or greedy methods) could be used to solve the right hand side equation of \eqref{equ:Samples1b} under the sparsity constraint $\|\bx[n]\|_{0} \leq N$ (see, e.g., \cite{FoucartRauhut_CS}).

The signal model implies that $\bx[n]$ has the same sparsity pattern for each $n\in\ZN$, i.e.
the index set $\mathcal{I} = \mathcal{I}(\bx) = \{k\ : x_{k}[n] \neq 0 \}$ is independent of $n$.
Consequently, it is sufficient to determine the sparsity pattern only once. 
Afterwards, one can solve \eqref{equ:Samples1b} easily using the knowledge of $\mathcal{I}(\bx)$.
Indeed, once $\mathcal{I}(\bx)$ is known, one can build the submatrix $\bB_{\mathcal{I}}$, which contains the columns of $\bB$ indexed by $\mathcal{I}(\bx)$.
Then, since $\kr(\bB) \geq 2N$, the matrix $\bB_{\mathcal{I}}$ has full column rank, and so we can solve 
\begin{equation*}
	\bx_{\mathcal{I}}[n] = \bB^{\dagger}_{\mathcal{I}}\, \bz[n]\,,
	\qquad n\in \ZN\;.
\end{equation*}
where $\bB^{\dagger}_{\mathcal{I}} = (\bB^{*}_{\mathcal{I}}\bB_{\mathcal{I}})^{-1}\bB^{*}_{\mathcal{I}}$ is the Moore-Penrose pseudoinverse of $\bB_{\mathcal{I}}$ and where the vector $\bx_{\mathcal{I}}[n] \in \CN^{N}$ contains only the nonzero entries of $\bx[n]$, which are indexed by $\mathcal{I}(\bx)$.
Since $\bz[n]$ is only known up to a global phase factor, it is clear that also $\bx_{\mathcal{I}}[n]$ can be determined only up to such a phase factor.

To identify the index set $\mathcal{I}(\bx)$, it is advisable to use more than one measurement vector $\bz[n]$.
For example, one can collect consecutive vectors $\bz[n]$ and $\bx[n]$ as columns of a matrix $\bZ = (\bz[N_{0}] , \dots, \bz[N_{1}])$ and $\bX = (\bx[N_{0}],\dots,\bx[N_{1}])$, respectively. Then the right hand side of \eqref{equ:Samples1b} becomes
\begin{equation}
\label{equ:MMV}
	\bZ = \bB\, \bX
\end{equation}
where the matrix $\bX$ has only $N$ nonzero rows. Such a sparse multiple measurement vector (MMV) problem is known to better identify a sparse model than a single measurement vector problem \cite{Cotter_IEEESP05,Tropp_SP06}. If $\kr(\bB) \geq 2 N$ then it is known that \eqref{equ:MMV} has a unique $N$-sparse solution $\bX$.
This solution will have only $N$ nonzero rows, which determine the index set $\mathcal{I}(\bx)$.

We note that there exist different and more systematic ways to obtain an MMV problem similar to \eqref{equ:MMV} from the data $\{\bz[n]\}$ and which guarantee a unique $N$-sparse solution \cite{ME_TheoryToPraxis10}.
Solving \eqref{equ:MMV} is again an NP-hard problem, but several suboptimal efficient algorithms for solving \eqref{equ:MMV} are known \cite{Cotter_IEEESP05,Tropp_SP06}.

\paragraph*{IV. Interpolation}

The $l$-th row of the vector sequence $\{ \bx[n] \}_{n\in\ZN}$ contains the signal samples $\{x_{l}[n]\}_{n\in\ZN}$ of the signal component from the $l$-th frequency band of length $\Omega$. So we can recover the signal $x\in \PW_{\B}$ by the usual Shannon sampling series
\begin{equation*}
	x(t) = \sum_{l\in\mathcal{I}(\bx)} \left( \sum_{n\in\ZN} x_{l}[n]\, \frac{\sin(\pi[\Omega t - n])}{\pi[\Omega t - n]} \right)\, \E^{-\I 2\pi l \Omega t}\;,
\end{equation*}
where the multiplication with the exponential functions shifts the signal components back on the correct band locations.
The above sum is known to converge in the norm of $\PW_{\B}$ and uniformly on $\RN$.

\paragraph*{Summary}

We showed that if the modulating functions in Fig.~\ref{fig:Modul1} and \ref{fig:Modul2} are properly chosen, then we are able to recover almost every signal from the signal space $\PW_{\B}$. We summarize this result as follows.

\begin{theorem}
Let $x \in \PW_{\B}$ be a signal which occupies $N$ out of $L$ frequency bands of length $\Omega$.
For $M\geq 2 N$ there exists modulating functions $p_{0}$ and $p_{k,m}$ where $k=1,\dots,4$, $m=1,\dots,M-1$ such that $x$ can be reconstructed (up to a phase factor) from phaseless measurements taken with the sampling system of Fig.~\ref{fig:Modul1} and \ref{fig:Modul2}.
\end{theorem}

\begin{remark}
We actually designed concrete modulating sequences $p_{k,m}$ and $p_{0}$ which allow signal recovery from the phaseless measurements. However, as we discuss below, there are many different possible modulating sequences which allow for signal recovery from phaseless samples.
\end{remark}

\begin{remark}
Actually, $x \in \PW_{\B}$ can only be recovered if the first entry $z_{1}[n]$ of the vector $\bz[n]$ in \eqref{equ:Samples1b} is nonzero for all $n\in\ZN$.
If this condition is not satisfied, the phase propagation described in step II breaks down.
However, the set of all signals in $\PW_{\B}$ for which $z_{1}[n] = 0$ for some $n\in\ZN$ is a very thin set in $\PW_{\B}$ \cite{PYB_JFAA14}, such that the proposed approach will recover ``almost every'' signal in $\PW_{\B}$.
\end{remark}

\section{Discussion}

\paragraph*{Sampling rate}

The Fourier transform $\widehat{x}$ of any signal $x \in \PW_{\B}$ is supported in the frequency domain on a set of Lebesgue measure $2 (N/L) f_{\mathrm{N}}$. 
If the support of $\widehat{x}$ would be known, then there exists a sampling scheme which allows to reconstruct $x$ from samples taken at an average rate which can be arbitrarily close to the Laudau rate $R_{\mathrm{L}} := 2(N/L) f_{\mathrm{N}}$.

The sampling scheme described above has $4 M$ parallel branches, with $M \geq 2 N$.
In each branch the squared amplitude of the signals are sampled at a rate $R_{0} = 1/T_{s} = \Omega = 2 f_{\mathrm{N}}/L$. So the overall average sampling rate of the proposed scheme is equal to
\begin{equation*}
R = 16\, \tfrac{N}{L} f_{\mathrm{N}} = 8\, R_{\mathrm{L}}\;.
\end{equation*}
So compared with the situations where we have access to amplitude and phase measurements and where we have knowledge of the band locations, we need an eight times higher sampling rate.
We lose a factor $2$ because the band locations are unknown \cite{ME_TheoryToPraxis10}, and we lose an additional factor $4$, because only amplitude samples are available \cite{PYB_STIP14}.

\paragraph*{Extensions and variants}

We emphasize that our main focus in this contribution lies on the general approach to include phaseless measurements into sub-Nyquist sampling systems.
Using the sub-Nyquist approach of \cite{ME_BlindMultipand_IEEESP09} the infinite-dimensional problem reduces to a finite-dimensional CS problem.
Then the phaseless measurements can be implemented similarly as in finite-dimensional phase retrieval problems.
The proposed methodology allows easily to use different algorithms for phase retrieval and compressive sampling.
In \eqref{equ:Samples1b}, the matrix $\bPsi$ is the measurement matrix for the phase retrieval step. Here basically any known method of phase retrieval with an appropriated $\bPsi$ can be used.
Similarly, matrix $\bB$ is a measurement matrix for the compressive sampling step to determine the occupied bands of the signal.
The particular choices for $\bPsi$ and $\bB$ in this paper are primarily to illustrate the general approach.
The concrete choice of $\bPsi$ and $\bB$ determine the mixing sequences $p_{k,m}$ of the sampling system.

There are many technical details \cite{ME_BlindMultipand_IEEESP09,ME_TheoryToPraxis10} concerning the sub-Nyquist sampling system which we omitted for simplicity of the presentation.
Similarly, a slightly simplified signal model was used were the signal bands have fixed positions. The generalization to a model with arbitrary band locations as in \cite{ME_TheoryToPraxis10} is straightforward.

\paragraph*{Phase propagation}
The sub-Nyquist sampling reduces the infinite-dimensional problem to many finite-dimensional problems.
The additional phase retrieval step makes it necessary to transfer phase information among these finite-dimensional problems.
To this end it was necessary to introduce the additional preprocessing step of Fig.~\ref{fig:Modul2}.
Up to now there seems to exist no other solution to connect these finite dimensional blocks.
Clearly, this phase propagation step has the potential problem that also errors are propagated from one block to the next block.
In future investigations, the influence of this phase propagation on the overall performance and stability of signal recovery has to be analyzed in detail.


\end{document}